\documentclass{Format_Files/WileyMSP-template}

\usepackage{float}
\usepackage{graphicx} 
\usepackage[margin=0.5in]{geometry}
\usepackage[dvipsnames]{xcolor}
\usepackage{amsmath}
\usepackage{hyperref} 
\usepackage{multirow}
\usepackage{newtxtext}         \usepackage{chemformula} 
\usepackage{booktabs}

\usepackage[normalem]{ulem}
\usepackage[textwidth=1.5cm,textsize=footnotesize]{todonotes}
\hypersetup{colorlinks, 
	linkcolor={blue!75!black!80!yellow},
	citecolor={blue!75!black!80!yellow}, 
	urlcolor={blue!75!black!80!yellow}
	}
\usepackage[cmintegrals]{newtxmath}
\usepackage{lipsum}

\newcommand{\BostonCollege}{Department of Physics, Boston College, Chestnut Hill, MA, USA}

\begin{document}
\title{Engineering Anomalously Large Electron Transport in Topological Semimetals}
\maketitle

\author{Vincent M. Plisson$^{1*}$,}
\author{Xiaohan Yao$^{1*}$,}
\author{Yaxian Wang$^{3}$,}
\author{George Varnavides$^{2}$,}
\author{Alexey Suslov$^{4}$,}
\author{David Graf$^{4}$,}
\author{Eun Sang Choi$^{4}$,}
\author{Hung-Yu Yang$^{1}$,}
\author{Yiping Wang$^{1}$,}
\author{Marisa Romanelli$^{1}$,}
\author{Grant McNamara$^{1}$,}
\author{Birender Singh$^{1}$,}
\author{Gregory T. McCandless$^{5}$,}
\author{Julia Y. Chan$^{5}$,}
\author{Prineha Narang$^{2}$,}
\author{Fazel Tafti$^{1}$,}
\author{Kenneth S. Burch$^{1}$}

\begin{itemize}
    \item[] \BostonCollege 
    \item[] College of Letters and Science, University of California Los Angeles, Los Angeles, California 90095, USA
    \item[] Institute of Physics, Chinese Academy of Sciences, Beijing 100190, China
    \item[] National High Magnetic Field Laboratory, Florida State University, Tallahassee, Florida 32310, USA
    \item[] Department of Chemistry and Biochemisty, Baylor University, Waco, TX, 76798, USA
\end{itemize}




\begin{abstract}
Anomalous transport of topological semimetals has generated significant interest for applications in optoelectronics, nanoscale devices, and interconnects. Understanding the origin of novel transport is crucial to engineering the desired material properties, yet their orders of magnitude higher transport than single-particle mobilities remain unexplained. This work demonstrates the dramatic mobility enhancements result from phonons primarily returning momentum to electrons due to phonon-electron dominating over phonon-phonon scattering. Proving this idea, proposed by Peierls in 1932, requires tuning electron and phonon dispersions without changing symmetry, topology, or disorder. This is achieved by combining de Haas - van Alphen (dHvA), electron transport, Raman scattering, and first-principles calculations in the topological semimetals MX$_2$ (M=Nb, Ta and X=Ge, Si). Replacing Ge with Si brings the transport mobilities from an order magnitude larger than single particle ones to nearly balanced. This occurs without changing the crystal structure or topology and with small differences in disorder or Fermi surface. Simultaneously, Raman scattering and first-principles calculations establish phonon-electron dominated scattering only in the MGe$_2$ compounds. Thus, this study proves that phonon-drag is crucial to the transport properties of topological semimetals and provides insight to further engineer these materials.
\end{abstract}

\begin{figure}
    \centering
    \includegraphics[width=1\textwidth]{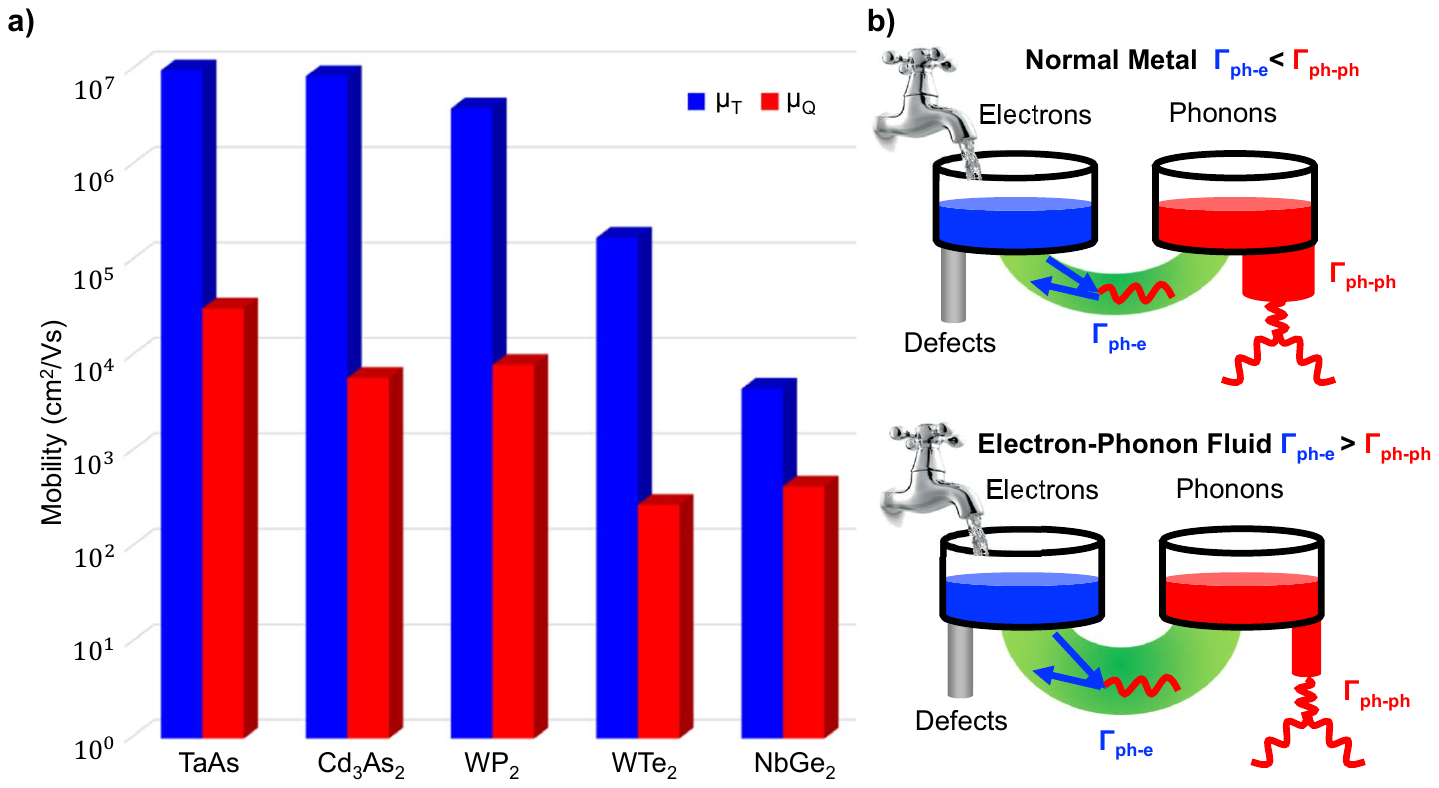}
    \vspace{-8ex}
    \caption{a) Transport (blue) and single particle (red) mobilities for various topological semimetals. Despite significant differences in topology, symmetry, and bulk band structures, these materials exhibit orders of magnitude higher transport than single particle mobilities.  b) Cartoon schematic of electron scattering pathways in a normal metal (top) and those in a phonon-electron fluid (bottom). The faucet represents an electric field providing energy and momentum to the electrons. The momentum of these electrons can be lost either by defect scattering or into the phonon bath. From here, anharmonic phonon decay transfers the energy and momentum into heat. However, if the phonon-electron scattering overwhelms the phonon-phonon, the momentum is instead returned to the electron bath, greatly enhancing bulk transport.}
    \label{fig:anomalous transport}\vspace{-6ex}
\end{figure}

Predictions and observations of novel electronic and optoelectronic responses have generated great interest in topological materials due to their potential for device applications and interconnects\cite{chen2020topological,ma2021topology,liu2020semimetals}. As shown in Figure \ref{fig:anomalous transport}a, a particularly perplexing aspect of the transport in topological semimetals is the three orders of magnitude larger transport ($\mu_T$) mobilities than their single particle  ($\mu_Q$) ones. Such enhancements could pave the way to future low-loss electronic interconnects,  but this is currently limited to low temperatures. Thus, there is an urgent need to uncover the origin of these extreme transport mobilities. Some potential explanations have been proposed, ranging from symmetry, topological protection, and chirality-protected backscattering\cite{doi:10.1146/annurev-matsci-070218-010023}. Nonetheless, experiments have not directly demonstrated how a specific material property, when removed, eliminates this behavior. 

Such experiments are crucial as the apparent universality of the anomalous transport in semimetals calls into question the previous potential explanations. For example, if the mechanism's origin results from topology, explaining materials such as \ch{WP2} where the Weyl node is over 300 meV below the Fermi surface is challenging. It is also unclear why such a mechanism would work in materials with different quantum geometries, as anomalous transport appears in Dirac (\ch{Cd3As2}), type-I (\ch{TaAs}), and type-II (\ch{WP2},\ch{WTe2}, \ch{NbGe2}) Weyl semimetals (see: Figure \ref{fig:anomalous transport}a). Furthermore, these materials possess different crystal structures, symmetries, and Fermi surfaces of widely varying sizes. However, some Raman scattering experiments suggest a common factor in a few of these materials: phonon-electron scattering exceeds phonon-phonon\cite{PhysRevX.11.011017,PhysRevB.102.075116,PhysRevB.98.115130,PhysRevB.94.094302, joshi2016phonon, PhysRevB.95.235148}.

Unlike normal metals, this scenario suggests topological semimetals are in the regime of phonon-electron fluids. Specifically, the phonons primarily decay via phonon-electron rather than phonon-phonon scattering.\cite{yangNComm2021,PhysRevX.11.011017} To see how this affects electron transport, let us review the fate of energy and momentum imparted by an electric field to the electrons. As illustrated in the top of Figure \ref{fig:anomalous transport}b, in a normal metal, devoid of Umklapp scattering, the excited electrons lose energy and momentum by scattering off defects and phonons\cite{PhysRevB.103.155128}. The momentum imparted to the phonons through these collisions is then lost to heat dissipation via anharmonic phonon-phonon scattering. However, as originally described by Peierls in 1932,\cite{peierls1932frage} if one can engineer the phonon-electron scattering rate ($\Gamma_{ph-e}$) to be greater than the phonon-phonon one ($\Gamma_{ph-ph}$), then the phonons would enhance the charge transport by returning the momentum to the electrons (bottom of Figure \ref{fig:anomalous transport} b). Such a mechanism was proposed to explain the resistivity of some simple metals\cite{PhysRev.143.393,danino1981phonon, takayama2020robust}; however, the phonon lifetimes and scattering mechanisms were never directly probed. Indeed, to properly investigate this possibility, one needs to measure the phonon scattering while systematically tuning the anharmonicity or phonon joint density of states. This must be done without large changes in the system's disorder or affecting its topology, symmetry, or spin-momentum locking. This has proven to be a significant challenge. For example, efforts focused on germanium and silicon alloys to tune between dominant phonon-phonon and phonon-electron scattering suffered from large changes in defect scattering\cite{bhandari1980silicon,settipalli2022effect}. 

The MX$_2$ family of materials provides a unique solution to test the role of phonon-electron scattering in electron transport with minimal changes in other material properties. Indeed, these materials are established type-II Weyl semimetals, with relatively large transport mobilities and evidence for a phonon-electron fluid in \ch{NbGe2}\cite{PhysRevB.102.235144,yangNComm2021}. Furthermore, recent calculations suggest that phonon-electron coupling differs dramatically between the germanides and silicides\cite{PhysRevMaterials.5.L091202}.  With this in mind, we carefully measured their single particle and transport mobilities. As shown in Figure \ref{fig:Transport}a, we find the MGe$_2$ compounds are similar to other topological semimetals with an order of magnitude difference between $\mu_T$ and $\mu_Q$ for NbGe$_2$. Upon moving to \ch{TaGe2}, we observe the transport mobility is reduced by a factor of two, despite the single particle mobility doubling. Nonetheless, \ch{TaGe2} has a similar trend to the Nb analog with $\mu_T \gg \mu_{Q}$, suggesting it is on the border between anomalous and normal metal transport. Upon moving to the MSi$_2$ compounds, we observe normal metallic transport with nearly equal quantum and transport mobilities. At first glance, this is rather surprising as all four compounds have nearly identical crystal structures, symmetry, and topology. Furthermore, as shown later, going from Ge to Si does not significantly change the Fermi surface size or disorder. However, as we reveal through measurements of the phonon scattering rates, Hall, quantum oscillations, and first-principles calculations, the change from germanium to silicon results in a switch from phonon-electron to phonon-phonon dominated scattering. This results from subtle changes in the acoustic mode bandwidth and electronic band structure near the Fermi level. We, therefore, provide strong evidence that phonons are at the heart of the remarkable transport behavior in semimetals.

\section*{Electronic Properties} 
\begin{figure}
    \centering
    \includegraphics[width=0.9\textwidth]{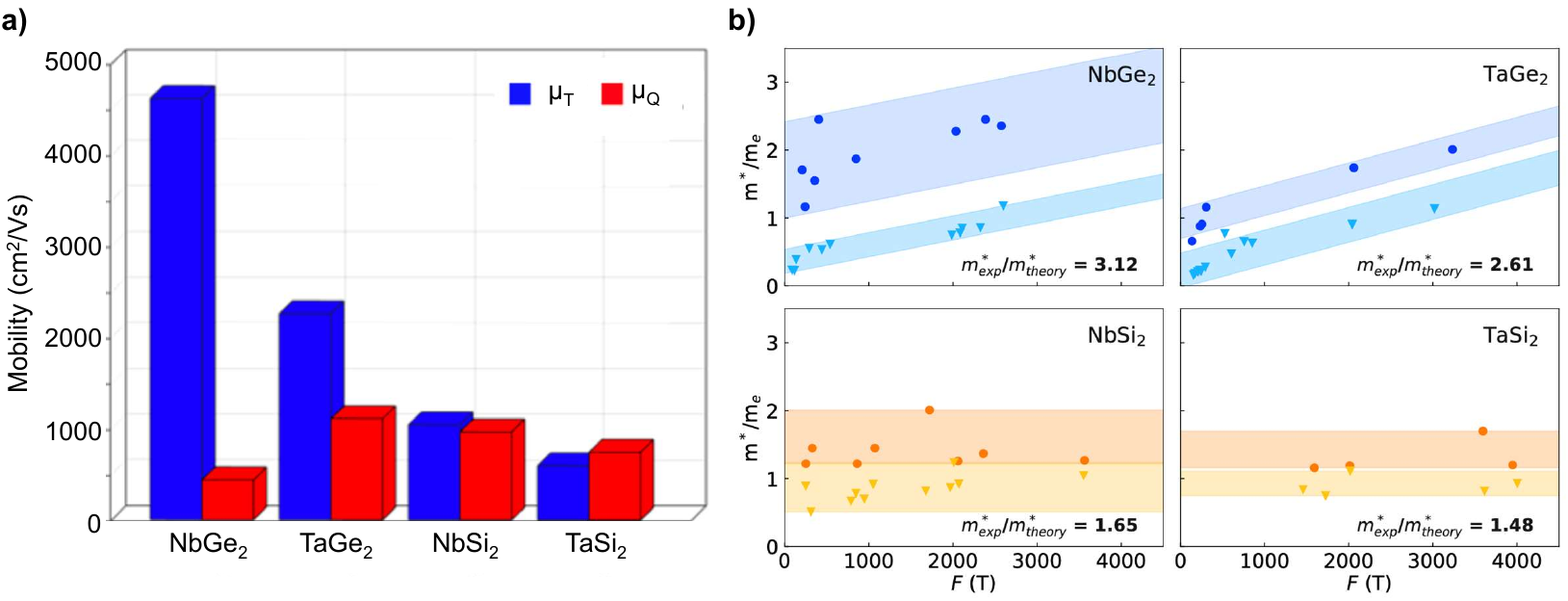}
    \caption{a) Single particle and transport mobility of all four compounds in the MX$_2$ family. As the series evolves from NbGe$_2$ to TaSi$_2$, a clear trend emerges with the transport mobility ($\mu_T$) decreasing dramatically despite larger single particle mobilities ($\mu_Q$). b) The experimental ($m^*_{exp}$) and theoretical ($m^*_{theory}$) quasiparticle effective masses plotted as a function of dHvA frequencies. A substantial enhancement in the experimental masses over the DFT calculated ones is observed in the MGe$_2$, but not MSi$_2$. Nonetheless, the overall change in mass between the materials does not explain the trends in mobilities.}
    \label{fig:Transport}
\end{figure}

We began our investigation by determining the electron mobilities for each material. Specifically, transport mobilities were extracted from the longitudinal and Hall resistivities. De Haas - van Alphen (dHvA) experiments were also performed to determine the single particle lifetimes. These experiments allowed us to investigate whether changes in the electronic properties alone can explain anomalous transport behavior ($\mu_T\gg \mu_Q$), specifically Fermi surface differences, carrier densities, or single particle mobilities (i.e., disorder). The longitudinal and transverse resistivity ($\rho_{xx}$ and $\rho_{xy}$) were measured from 2 K to 300 K with magnetic fields from 0 to 9 T. The full data sets at 2 K are shown in Supplemental Figure S1. The transport mobility, $\mu_T$, was obtained through a three-band model fit to both $\rho_{xx}$ and $\rho_{xy}$ simultaneously according to the following expressions\cite{yao2023large,PhysRevB.96.235128}
\begin{eqnarray}
\label{eq:multi}
\sigma_{xx}(H)=\frac{\rho_{xx}(H)}{\rho_{xx}^2(H)+\rho_{xy}^2(H)}=\sum_i \frac{\sigma_i}{1+\mu_i^2H^2}, \\
\sigma_{xy}(H)=\frac{\rho_{xy}(H)}{\rho_{xx}^2(H)+\rho_{xy}^2(H)}=\sum_i \frac{\sigma_i\mu_iH}{1+\mu_i^2H^2},
\end{eqnarray}
where the conductivity of each band ($\sigma_i$) is related to its transport mobility ($\mu_i$) through $\sigma_i=n_iq_i\mu_i$ with $q_i=\pm e$ for a hole or electron band. Adding more bands to the model did not produce additional meaningful information. For the results shown in Figure \ref{fig:Transport}a, we used an average of all $\mu_i$ to obtain the transport mobility $\mu_T$ for each \ch{MX2} compound.

 To ascertain the single-particle mobility $\mu_Q$ for each \ch{MX2} compound, we turned to quantum oscillations probed via dHvA. Here, we measured the magnetic torque in each \ch{MX2} sample subjected to a magnetic field from 0 to 41 T at $T=300$ mK. All four compounds showed rich patterns of oscillations with several frequencies, each corresponding to one branch of the Fermi surface. Using a detailed analysis shown in the Supplementary Information (Figure S1 and S2), we extracted the relaxation time on each Fermi surface $\tau_j$ and determined quantum mobility through $\mu_{j}=q_j \tau_j/m^*_j$ with $q_j$ and $m^*_j$ being the sign and effective mass of carriers on each cyclotron orbit \cite{gaudet2021weyl}. The individual quantum mobilities for each branch of the Fermi surface are tabulated in Supplementary Tables S1 through S4. Similar to the transport mobility $\mu_T$, the average of these mobilities was taken as $\mu_Q$ in the histograms of Figure \ref{fig:Transport}a.

As described earlier, the results of these experiments demonstrate the germanides reveal anomalous transport ($\mu_T\gg \mu_Q$) mobilities, while the silicides have nearly normal metallic responses ($\mu_T \approx \mu_Q$). Thus, we carefully evaluated the quantum oscillation data with our first-principles calculations. Specifically in Figure \ref{fig:Transport}b, we plot the experimentally determined and first-principles calculated effective masses versus the dHvA frequencies for each orbit observed. We find excellent agreement between the calculated and measured frequencies for all four materials. As shown in Supplementary Figure S4, while the calculated Fermi surfaces are very similar, there are some slight differences between the silicides and germanides. Specifically, the separate pockets in the first two bands in the silicides are absent in the germanides. However, these small differences cannot consistently explain the trend in the mobilities. The experimentally determined single particle mobilities are relatively similar between \ch{TaGe2}, \ch{NbSi2}, and \ch{TaSi2}. This is in spite of the fact that the transport mobility for \ch{TaGe2} is nearly double that found in the silicides, consistent with a much cleaner sample. In addition, the $\mu_{Q}$ for \ch{NbGe2} is far lower than the rest, yet it possesses a far greater transport mobility. This indicates changes in disorder, and the size of the Fermi surface cannot explain the drastic changes between the germanides and silicides. 

Returning to the mobilities, we further investigated the role of the effective mass of the carriers and, thus, the density of states at the Fermi level. Similar to our previous study of \ch{NbGe2}\cite{yangNComm2021}, we found that \ch{TaGe2} also shows an enhancement in the experimentally measured effective mass $m^*_{exp}$ compared with the DFT theoretical values $m^*_{theory}$. Since these materials possess weak electronic correlations and are non-magnetic, phonon-electron interactions are the prime suspects for the observed mass enhancement. Interestingly, this enhancement is largely suppressed in the silicides, consistent with their transport behavior returning to that of a normal metal.  

To see whether the effective masses can explain the electronic transport ($\mu_T$), let us quantitatively examine the differences between the germanides and silicides. Starting with the evolution from \ch{NbGe2} to \ch{TaGe2}, we note a two-fold increase in the single particle mobility $\mu_Q$. This occurs alongside a similar size reduction in the effective masses for nearly all orbits. Effective masses nearly half that of \ch{NbGe2} are also seen in the silicides. In addition, as mentioned earlier, the silicides reveal nearly identical single particle mobilities as \ch{TaGe2}. Thus, the changes in single particle mobilities across the MX$_{2}$ series can primarily be attributed to the effective mass. This is consistent with nearly similar levels of disorder for all four compounds. Nonetheless, the transport mobilities $\mu_T$ in \ch{MSi2} are also reduced by a factor of two from those found in \ch{TaGe2}. Taken together, it appears a consistent explanation for the transport and single particle mobilities cannot be found in the disorder or changes in the electronic structure alone. These high transport mobilities that are insensitive to the single particle scattering also imply dominant forward scattering or processes where the momentum from collisions is returned to the electrons. 

\section*{Phonon Scattering Mechanisms}

\begin{figure}
    \centering
    \includegraphics[width=0.9\textwidth]{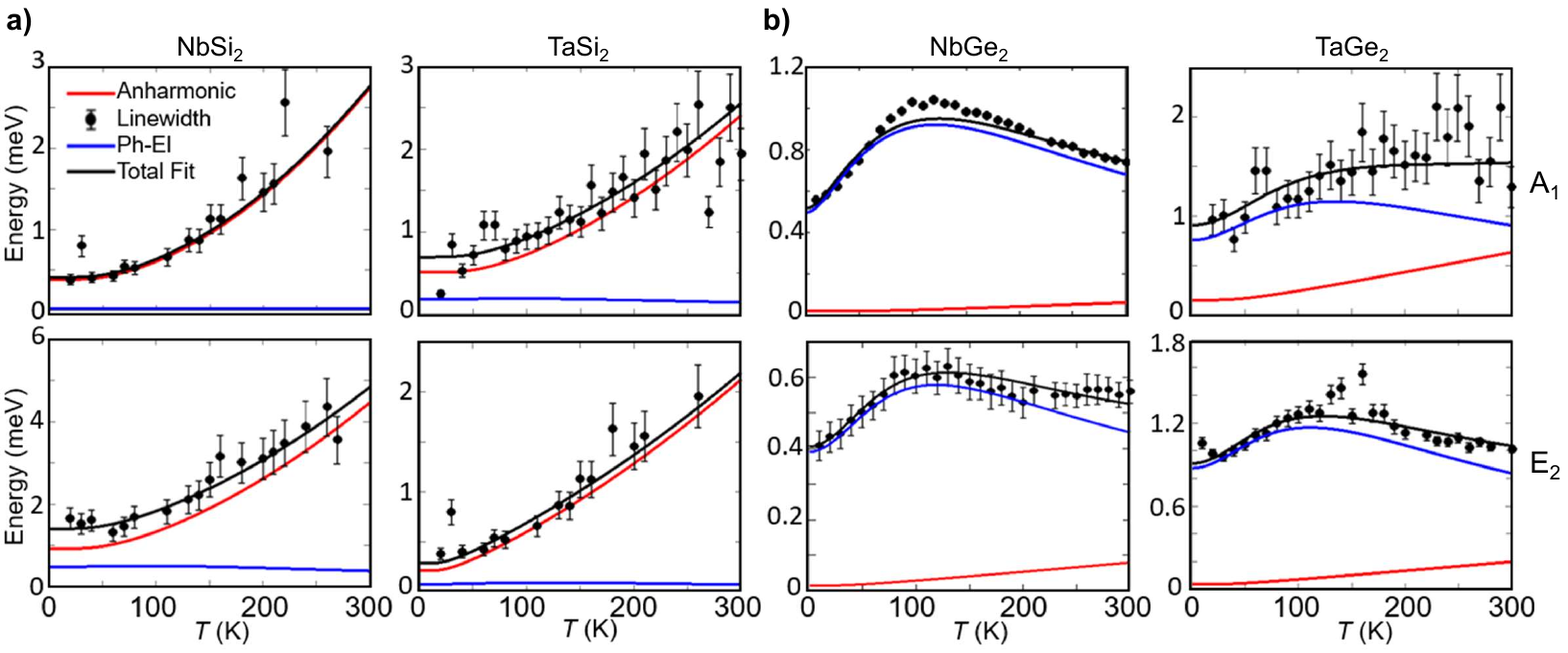}
    \caption {Temperature dependence of the Raman linewidths for A$_1$ and E$_2$ modes in a) MSi$_2$ and b) MGe$_2$. 
    Measured data are shown in black, the blue curve is a fit to the phonon-electron decay model, while the red curve is the typical anharmonic phonon-phonon contribution. The black line is a sum of both contributions and gives the total fit. We see clear evidence of phonon-phonon dominated linewidths in \ch{MSi2} and phonon-electron dominated linewidths in \ch{MGe2}.}
    \label{fig:Raman}
\end{figure}

To explore the role of phonons in the anomalous electronic transport of the \ch{MX2} series, we turned to Raman spectroscopy. This tool is excellent for measuring phonon frequencies, line widths, and symmetries versus temperature. With this information, we can uncover subtle changes in the lattice structure, anharmonicity, and dominant scattering pathways for each material\cite{tian2017understanding,yoon2011negative,riccardi2019probing}. As such, Raman spectroscopy provides a high-resolution probe into the behavior of the optical modes and has already been used to investigate novel states and phonon-electron coupling in topological materials\cite{PhysRevB.102.075116,PhysRevX.11.011017,PhysRevB.100.201107,PhysRevB.95.235148,shojaei2021raman,PhysRevB.100.201107,wang2022axial}. By careful analysis of the temperature dependence of the Raman linewidths, we can determine the decay mechanisms involved for the optical phonons in each compound. Here, we rely on the distinct temperature dependencies of anharmonic, defect, and phonon-electron scattering (Figure \ref{fig:Raman}). 
\\
In the case of anharmonic phonon-phonon scattering, we consider the decay of an optical phonon into two or more acoustic modes, illustrated in Figure \ref{fig:Transport and Linewidths}a. The Klemens model describes the resulting temperature dependence of the linewidth.\cite{han1993anharmonic} It is characterized by a monotonically decreasing linewidth with reduced temperature that gives way to low-temperature saturation due to the phonon's Bose statistics. As such, the anharmonic decay linewidth ($\Gamma_{Anharm}$) is typically modeled by a Bose-Einstein distribution ($n_B$) function at half (or one-third) of the frequency of the original mode (to account for energy and momentum conservation)
\begin{equation}
\label{eqn: Klemens}
    \Gamma_{Anharm}(T) = A \left( 1 + 2 n_{B}(\omega_{0}/2,T) \right) + B \left( 1 + 3 n_{B}(\omega_{0}/3,T) + 3 (n_{B}(\omega_{0}/3,T))^{2} \right).
\end{equation}

To this end, we measured the Raman response of the entire \ch{MX2} family from 10 K to 300 K with temperature steps of 10 K. The linewidth versus temperature extracted from fitting the spectra are shown in Figure \ref{fig:Raman} for representative modes (see the Supplemental section for details of the fitting procedures). Focusing first on the MSi$_2$ compounds shown in Figure \ref{fig:Raman}a, we see nearly temperature-independent linewidths from 10 K to 50 K. This is followed by a monotonically increasing linewidth that is approximately tripled by 300 K. This temperature dependence is a typical signature of anharmonic phonon-phonon scattering and is primarily described by the Klemens model. 

A qualitatively different behavior is observed in the MGe$_2$ compounds (Figure \ref{fig:Raman}b). Here, the linewidths begin to broaden almost immediately with increasing temperature, reaching a maximum value of one and a half to two times larger than the low-temperature linewidths. This is followed by a high-temperature saturation or a decreasing linewidth, sometimes reduced back to its low-temperature value. As the Klemens model of anharmonic phonon-phonon decay follows a Bose-Einstein temperature dependence, it fails to explain linewidths that decrease with increasing temperature. This implies that there are scattering pathways other than anharmonic phonon-phonon decay. Indeed, these follow the established pattern of dominant phonon-electron scattering in semimetals\cite{PhysRevX.11.011017,PhysRevB.95.235148,yangNComm2021,xu2017temperature}. In this case, as illustrated in Figure \ref{fig:Transport and Linewidths}b, the optical phonons decay into an electron-hole excitation (i.e., a pair of Fermions). As such, the probability of scattering is sensitive to the occupation of the initial and final state of the electron involved. We therefore model the temperature dependence of phonon linewidths ($\Gamma_{ph-e}$) dominated by phonon-electron scattering by the difference between the Fermi functions ($n_F$) of the electron and hole following \cite{yangNComm2021,PhysRevX.11.011017}
\begin{equation}
\label{eqn: Ph-El}
    \Gamma_{ph-e}(T) = C\int\limits_{-Inf}^{Inf} d\omega_{A}[n_{F}(\hbar\omega_{A})-n_{F}(\hbar\omega_{A}+\hbar\omega_{0})]e^{\frac{-(\omega_{A}-\mu)^2}{2\sigma^2}}.
\end{equation} 
Here, $\omega_0$ is the phonon energy, and the difference in the two Fermi functions represents the decay into an electron-hole pair, while the quantity $\omega_A$ represents the separation of the lowest band to the Fermi level. In this work, we added a subtle yet important detail to the model, namely using a Gaussian distribution about $\omega_A$. Simply taking the difference between two Fermi functions implies that there is only one location in the electronic band structure where the optical phonon can decay into an electron-hole excitation. By introducing a Gaussian distribution about $\omega_A$ and writing the fit as a convolution between this term and the difference in the Fermi functions of the electron and hole, we can account for multiple allowable excitations (Figure \ref{fig:Transport and Linewidths}b) and capture the entire 3D Fermi surface. Indeed, we found this improved model more accurately describes the range of temperature-dependent phonon lifetimes we observe.
 
 The qualitative nature of phonon-electron-dominated linewidths is explained in the following manner. The initial state must first be thermally populated for the electron to be excited to a different band. As the temperature increases, the lower band has a larger thermal population of electrons, leading to an increasing linewidth. Eventually, the states in the upper band begin to get filled while the lower band is emptied. This leads to high-temperature saturation and a decreasing linewidth as the interband transition becomes increasingly Pauli-blocked. A special case occurs when the initial state of the electron is at the Fermi level. There, at base temperature, the lower band is fully occupied, such that temperature will lower the probability of the phonon-electron decay. This produces a maximum for the linewidths at low temperatures that decreases with increasing temperature, as is seen for the $E_{2}(4)$ phonon mode in NbGe$_2$ \cite{yangNComm2021}.
\\
Before turning to the implications of our findings, we first consider an alternative explanation for the observed anomalous linewidth behavior in the germanides. As shown in the first panel of Figure \ref{fig:Raman}b, below 100 K, it is possible to fit the \ch{MGe2} linewidths to a model of anharmonic decay, as indicated by the black dashed line. This fit would suggest the Ge compounds have stronger anharmonic scattering than their Si counterparts. This is consistent with what would be expected for a heavier atom. The sudden change in the linewidths in the $100~K\rightarrow 150~K$ temperature range might then result from a dramatic change in the lattice at these temperatures. However, such a structural change is straightforwardly excluded from our X-ray diffraction and Raman measurements (See supplemental Figure S3 and Table S5). 

Alternatively, this effect could result from a dramatic shift in the lattice expansion as a function of temperature. Indeed, such effects have been reported in graphene\cite{yoon2011negative}, although there was no indication that the linewidths deviate from anharmonic decay as seen in the germanide compounds (see Supplemental Materials). Instead, graphene has an inflection point in the energy shifts of the optical phonons due to a negative thermal expansion coefficient. As shown in the supplemental Figure S3, we found that in the MGe$_2$ and MSi$_2$ compounds, the energy shift of every phonon follows typical anharmonic behavior with no unusual characteristics. This implies that the behavior of the phonon linewidths does not originate from changes in the lattice.  Furthermore, X-ray diffraction data reveal the lattice constants for all four compounds change by at most 0.5\% (Supplemental table S5) in this range, decreasing with temperature as expected.

\section*{Impact of Phonon Dynamics on Transport}

\begin{figure}[H]
    \centering
    \includegraphics[width=1\textwidth]{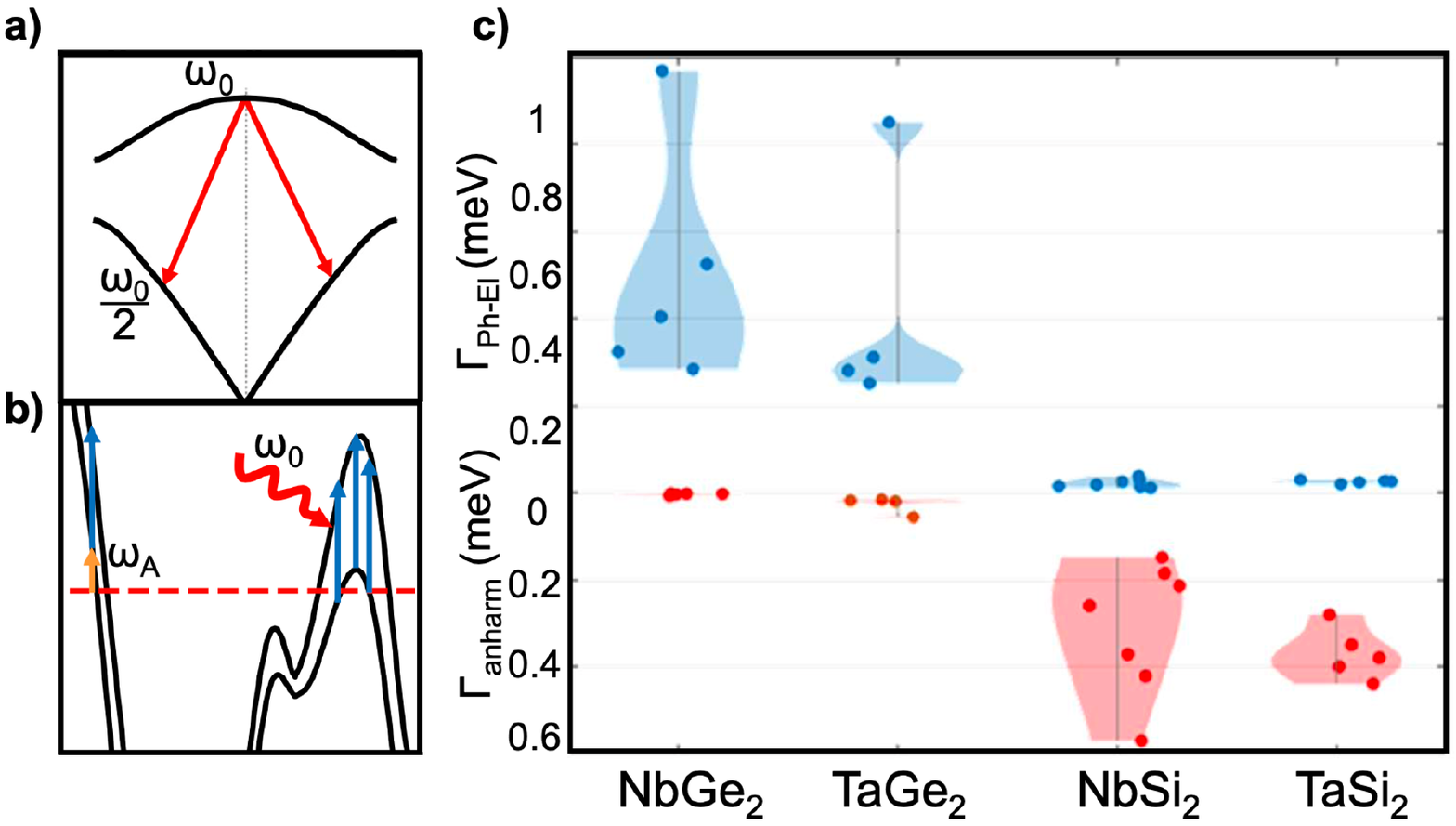}
    \caption {a)Cartoon schematic of anharmonic decay of an optical phonon into a pair of acoustic phonons, conserving energy and momentum.  b) Schematics of phonon decay into an electron-hole pair plotted on the bandstructure of \ch{NbGe2}. $\omega_0$ is the phonon energy, the orange arrow is $\omega_A$ and represents the separation of the band from the Fermi level. The blue arrow represents an electron-hole excitation of the same energy as the incident phonon.
    c) Phonon-electron and anharmonic contributions to Raman linewidths of every phonon mode in NbGe$_2$, TaGe$_2$, NbSi$_2$ and TaSi$_2$. The phonon-electron contribution heavily dominates the \ch{MGe2} linewidths, while the phonon-phonon contribution dominates the \ch{MSi2} linewidths.}
    \label{fig:Transport and Linewidths}
\end{figure}

To quantify the difference in phonon scattering rates between the compounds, we fit the linewidths to the sum of Equation~\ref{eqn: Klemens} and Equation~\ref{eqn: Ph-El}. This accounts for the possibility of the optical modes decaying into both electrons and phonons. The fits for each mode were obtained by first assuming a contribution from both scattering processes. To be conservative in our estimation of the scattering rates, we also assumed that linewidths showing characteristics of phonon-electron decay had as large of a contribution as possible from anharmonic scattering and vice-versa. That is to say, for the germanides, we first start by fitting to the Klemens model and gradually reduce the strength of the anharmonic scattering as we increase the contribution from the phonon-electron coupling. We followed the same process for modes that show typical anharmonic decay but in the opposite direction. This analysis allows us to quantify the coupling strengths for each mode without assuming that there is only one decay process at play, and it was done for every resolvable phonon mode in each material.

Having established the nature of the phonon scattering in the MX$_2$ family, we now attempt to correlate the phonon scattering with the anomalous transport in NbGe$_2$ and TaGe$_2$. To do so, we summarize the Raman results in Figure \ref{fig:Transport and Linewidths}c. Here, we plot the low-temperature Raman linewidths of all four of the MX$_2$ compounds with the contribution from phonon-electron scattering in the top half of the plot and the anharmonic scattering contribution in the lower half of the plot. From the Raman (Figure \ref{fig:Transport and Linewidths}c) and transport (Figure \ref{fig:anomalous transport}a) data for the MX$_2$ family, we see the phonon-electron fluid vanishing as the material is tuned from germanium to silicon. Raman linewidths of \ch{NbGe2} revealed the strongest relative phonon-electron coupling, and correspondingly, it has the largest bulk transport mobility. As we move to TaGe$_2$ we start to see that the phonon-electron fluid is on the verge of disappearing. Indeed, while $\Gamma_{ph-e}>\Gamma_{Anharm}$, for the average phonon-electron scattering rate across all modes, the ratio is lower in \ch{TaGe2} than \ch{NbGe2}. The phonon-phonon scattering becomes dominant upon replacing the germanium with silicon, and the phonon-electron fluid vanishes. Since the phonons no longer return momentum to the electron bath, the transport mobility is reduced by half an order of magnitude as we return to a normal metal.

The transport and Raman results clearly reveal a correlation between the dominant phonon-electron scattering and high transport mobility in NbGe$_2$ and TaGe$_2$. This enhancement of transport when phonon-electron rather than phonon-phonon pathways dominate phonon scattering is consistent with the phonon-electron fluid first described by Peierls in 1932\cite{peierls1932frage}. To the best of our knowledge, this is the first direct demonstration that removing only the dominance of phonon-electron scattering results in a dramatic reduction in transport mobilities.

\section*{Tipping the balance between phonon-phonon and phonon-electron}
\begin{figure}
    \centering
    \includegraphics[width=0.9\textwidth]{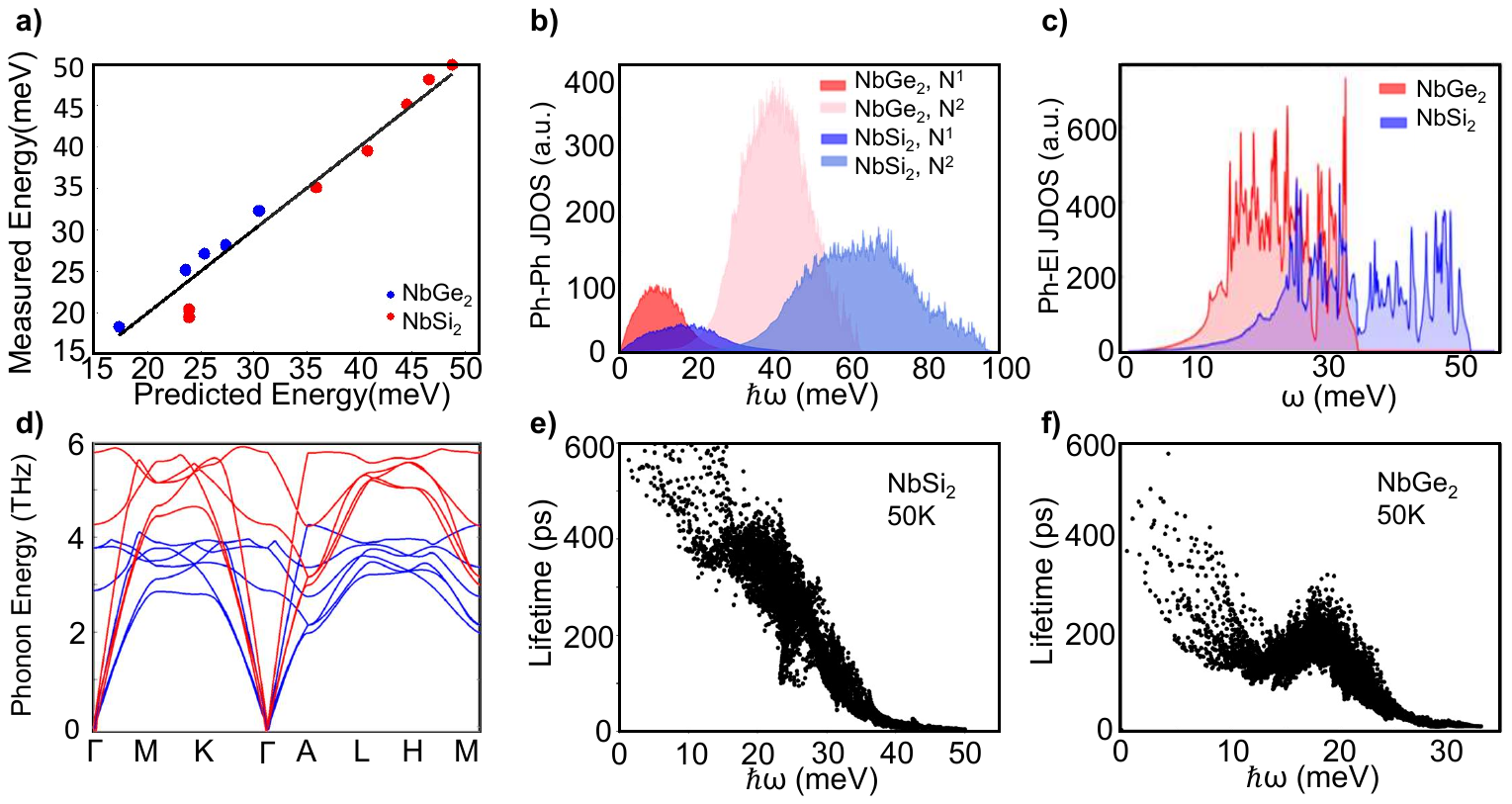}
    \caption{a) Calculated vs. measured phonon frequencies for \ch{NbGe2} and \ch{NbSi2}. The black line indicates the ideal agreement. b) Phonon-phonon joint density of states (JDOS) for NbGe$_2$ and NbSi$_2$ which show nearly equal spectral weight despite reduced acoustic bandwidth in \ch{NbSi2} c) Phonon-electron JDOS for NbGe$_2$ and NbSi$_2$ which shows approximately 20\% larger phase space for phonon-electron decay in NbGe$_2$ than NbSi$_2$. d) Phonon dispersions of NbGe$_2$ and NbSi$_2$ are shown in blue and red, respectively. Dispersions show approximately double the acoustic bandwidth for NbSi$_2$ than NbGe$_2$. e) Calculated phonon-phonon lifetimes at 50K for \ch{NbGe2} and \ch{NbSi2}, respectively.}
    \label{fig:DFT}
\end{figure}

We turn to the electron and phonon dispersions to understand the absence of a phonon-electron fluid in the silicides. As discussed earlier, our first-principles calculations are in good agreement with the experimental dHvA results. Furthermore, the electronic band dispersion and Fermi surfaces alone cannot explain the difference in transport between the silicides and germanides. This is consistent with previous works indicating that the MGe$_2$ and MSi$_2$ compounds have similar electronic band dispersion~\cite{PhysRevMaterials.5.L091202}. 
\\
We first compared our measured phonon energies to those calculated from first principles to validate our calculations. The results of this comparison are shown in Figure \ref{fig:DFT}a, where we see a good agreement for \ch{NbGe2} and \ch{NbSi2}. With confidence that first principles accurately capture these materials, we calculate the acoustic phonon dispersions since they govern the low-temperature, phonon-mediated electronic transport. The resulting dispersions are shown in Figure \ref{fig:DFT}d. We find that the acoustic phonons extend to roughly 12 meV in NbGe$_2$, and to 20 meV in NbSi$_2$ due to its lighter group 14 element. We calculated the phase space for both materials to determine if the enhanced bandwidth is responsible for greater phonon-phonon scattering. We considered two types of scattering events: up conversion and down conversion. Namely, one phonon mode can decay into two lower-frequency branches or interact with another phonon branch and generate a mode with a higher frequency. They are evaluated via:
\begin{equation}
 N^1(q,\omega) = \frac{1}{N}\sum_{\lambda \lambda'} \Delta (-q+q'+q'')\left[ \delta(\omega +\omega_\lambda'-\omega_\lambda'')-\delta(\omega - \omega_\lambda' +\omega_\lambda'')\right] ;
\end{equation}
\begin{equation}
 N^2(q,\omega) = \frac{1}{N}\sum_{\lambda \lambda'} \Delta (-q+q'+q'')\delta(\omega -\omega_\lambda'-\omega_\lambda'').
\end{equation}
Here, $\Delta (-q+q'+q'')=1$ if $-q+q'+q''=k$ (the reciprocal lattice vector), and otherwise $\Delta (-q+q'+q'')=0$. Figure \ref{fig:DFT}b shows the two joint densities of states (JDOS) for NbGe$_2$ and NbSi$_2$. Intriguingly, despite the relatively higher acoustic bandwidth in NbSi$_2$ (Figure~\ref{fig:DFT}b), we find the integrated JDOS is nearly similar to that of \ch{NbGe2}. This may result from a phonon continuum in these materials. Indeed, unlike other compounds with an optical gap, the acoustic modes continuously evolve into the optical ones. Next, we calculated the phonon lifetimes resulting from phonon-phonon scattering. The results at 50 K for \ch{NbSi2} and \ch{NbGe2} are shown in (Figure \ref{fig:DFT}e and f), with the 300 K results shown in the supplemental. Consistent with our JDOS calculations, at low energies relevant to low-temperature transport, we find that the phonon-phonon lifetimes for \ch{NbGe2} and \ch{NbSi2} are nearly equal.

This surprising result motivated us to calculate the phase space for phonon-electron scattering following
\begin{equation}
     N(\omega) = \frac{g_s^2}{N_k N_k'}g(\mu)^2 \sum_{k,k',b,b',\alpha}\delta(\hbar\omega-\hbar\omega^\alpha_{k-k'})\delta(\mu-\varepsilon_k^b)\delta(\mu-\varepsilon_k'^b),
 \end{equation}
 where $\varepsilon_{k,b}$ denotes the electronic state with band index $b$ and wavevector $k$, $\omega$ the phonon frequency with mode index $\alpha$, and the prefactors are employed for the single-particle DOS renormalization. Figure \ref{fig:DFT}c shows the ph-e scattering phase space for both NbGe$_2$ and NbSi$_2$, as a function of phonon frequency. For both compounds, the scattering phase space for all-optical phonon modes is significant and distributed relatively evenly. While there is no drastic disparity between the two compounds, integrating over the phonon frequency yields roughly $\sim 20\%$ larger phase space for phonon-electron scattering in NbGe$_2$. This larger phase space for phonon-electron scattering also dramatically reduces phonon lifetimes in \ch{NbGe2}. Specifically, consistent with previous reports,\cite{PhysRevMaterials.5.L091202} we find the low energy phonon-electron lifetimes at 300 K in \ch{NbGe2} are $\approx 0.1$ ps, whereas in \ch{NbSi2} they are $\approx 10$ ps. This should be compared with the phonon-phonon lifetimes for both compounds at 300K (shown in supplemental Figure S6), which are 10 picoseconds. Thus, our calculations show the scattering rates of the acoustic modes have a ratio of $\tau_{ph-el}/\tau_{ph-ph}\approx10^{-1}$ for \ch{NbGe2}, while it is nearly unity for \ch{NbSi2}. 
 
Hence, our calculations confirm our experimental findings that phonon-electron scattering dominates in the germanides, while the silicides reside in the normal regime of phonon-phonon scattering. Furthermore, this occurs alongside the change from anomalous to normal metal transport without changes in topology, symmetry, or lattice structure and with nearly similar disorder and Fermi surface. We have, therefore, established that removing the dominant phonon-electron scattering alone results in the removal of anomalous transport mobilities. Hence, phonons are central to the enormous electron transport in topological semimetals. Our study also reveals that the dramatic difference between \ch{NbGe2} and \ch{NbSi2} cannot easily be attributed to the electronic or phonon properties alone. Indeed, the enhanced phonon-electron scattering in the germanides results from subtle details of both the electronic and lattice dispersions. Specifically, an electronic excitation near the Fermi level is only possible if the phonon energy and momentum match that needed to excite an electron-hole pair.

\section*{ Moving Forward}

 Though the quantum geometry of the electronic structure is not a contributing factor in the \ch{MX2} series, the topological semimetal state appears to be a common host for a coupled phonon-electron fluid. It seems the recipe that is critical for forming a phonon-electron fluid in these topological semimetals includes both: 1) heavy atoms with strong spin-orbit coupling and 2) a second atom with extended orbitals. However, it seems the light atom must be selected carefully to maintain the delicate balance between the phonon and electron dispersions. Indeed, for the MX$_2$ series, the silicon atom in TaSi$_2$ and NbSi$_2$ still enables a topological semimetal but slightly enhances the phonon-phonon scattering phase space while also dramatically reducing the overlaps in the electronic and lattice dispersions. Since NbGe$_2$ and TaGe$_2$ are already very close to the "tipping point," as soon as the relative phase spaces for scattering invert, the phonon-electron fluid disappears with a dramatic reduction in the transport mobility.
 
 Having an accurate understanding of the origin of the unique transport behavior in topological semimetals allows us to exploit the material properties that bring us into this transport regime. For example, if the acoustic bandwidth can be further reduced, this would prevent the growth of anharmonic scattering at high temperatures. As such, the phonon-electron fluid, and thus, large transport mobilities, could be extended to higher temperatures. Similarly, given the large difference in mean free paths between electrons and phonons, it is unclear how the phonon-electron fluid will be modified when these systems are reduced to the nanoscale. Indeed, these materials were shown to be promising candidates for electronic interconnects\cite{gall2021materials}. The strong enhancement of electron transport via phonons in these materials also suggests the phonon-electron fluid could be exploited via coherent phonon pumping to create novel states of matter\cite{ma2021topology,disa2021engineering}. Furthermore, a natural conclusion of our work is that the quantum geometry near the Fermi surface is not central to the anomalous transport in topological semimetals. This is a surprising result, especially given recent calculations suggesting quantum geometry is crucial for understanding enhancements in phonon-electron coupling\cite{PhysRevMaterials.5.L091202,yu2023nontrivial}. There have also been reports that phonons can induce topological transitions in Dirac materials\cite{garate2013phonon}. Nonetheless, further work is needed to see if the intricate interplay between the phonons and electrons in these systems means materials with topological phonons also reveal novel transport behavior.\cite{xu2022catalogue,PhysRevResearch.4.043060}

\newpage
\bibliography{Advanced_Mat_Submission}

\begin{thebibliography}{10}
\providecommand{\url}[1]{\texttt{#1}}
\providecommand{\urlprefix}{URL }

\bibitem{chen2020topological}
C.-T. Chen, U.~Bajpai, N.~A. Lanzillo, C.-H. Hsu, H.~Lin, G.~Liang,
\newblock In \emph{2020 IEEE International Electron Devices Meeting (IEDM)}. IEEE, \textbf{2020} 32--4.

\bibitem{ma2021topology}
Q.~Ma, A.~G. Grushin, K.~S. Burch,
\newblock \emph{Nature materials} \textbf{2021}, \emph{20}, 12 1601.

\bibitem{liu2020semimetals}
J.~Liu, F.~Xia, D.~Xiao, F.~J. Garcia~de Abajo, D.~Sun,
\newblock \emph{Nature materials} \textbf{2020}, \emph{19}, 8 830.

\bibitem{doi:10.1146/annurev-matsci-070218-010023}
J.~Hu, S.-Y. Xu, N.~Ni, Z.~Mao,
\newblock \emph{Annual Review of Materials Research} \textbf{2019}, \emph{49}, 1 207.

\bibitem{PhysRevX.11.011017}
G.~B. Osterhoudt, Y.~Wang, C.~A.~C. Garcia, V.~M. Plisson, J.~Gooth, C.~Felser, P.~Narang, K.~S. Burch,
\newblock \emph{Phys. Rev. X} \textbf{2021}, \emph{11} 011017.

\bibitem{PhysRevB.102.075116}
D.~Wulferding, P.~Lemmens, F.~B\"uscher, D.~Schmeltzer, C.~Felser, C.~Shekhar,
\newblock \emph{Phys. Rev. B} \textbf{2020}, \emph{102} 075116.

\bibitem{PhysRevB.98.115130}
J.~Coulter, R.~Sundararaman, P.~Narang,
\newblock \emph{Phys. Rev. B} \textbf{2018}, \emph{98} 115130.

\bibitem{PhysRevB.94.094302}
F.~Jin, X.~Ma, P.~Guo, C.~Yi, L.~Wang, Y.~Wang, Q.~Yu, J.~Sheng, A.~Zhang, J.~Ji, Y.~Tian, K.~Liu, Y.~Shi, T.~Xia, Q.~Zhang,
\newblock \emph{Phys. Rev. B} \textbf{2016}, \emph{94} 094302.

\bibitem{joshi2016phonon}
J.~Joshi, I.~R. Stone, R.~Beams, S.~Krylyuk, I.~Kalish, A.~V. Davydov, P.~M. Vora,
\newblock \emph{Applied physics letters} \textbf{2016}, \emph{109}, 3.

\bibitem{PhysRevB.95.235148}
A.~Sharafeev, V.~Gnezdilov, R.~Sankar, F.~C. Chou, P.~Lemmens,
\newblock \emph{Phys. Rev. B} \textbf{2017}, \emph{95} 235148.

\bibitem{yangNComm2021}
H.-Y. Yang, X.~Yao, V.~Plisson, K.~S. Burch, F.~Tafti,
\newblock \emph{Nat. Commun.} \textbf{2021}, \emph{12}, 1 5292.

\bibitem{PhysRevB.103.155128}
X.~Huang, A.~Lucas,
\newblock \emph{Phys. Rev. B} \textbf{2021}, \emph{103} 155128.

\bibitem{peierls1932frage}
R.~Peierls,
\newblock \emph{Annalen der Physik} \textbf{1932}, \emph{404}, 2 154.

\bibitem{PhysRev.143.393}
N.~Wiser,
\newblock \emph{Phys. Rev.} \textbf{1966}, \emph{143} 393.

\bibitem{danino1981phonon}
M.~Danino, M.~Kaveh, N.~Wiser,
\newblock \emph{Journal of Physics F: Metal Physics} \textbf{1981}, \emph{11}, 12 2563.

\bibitem{takayama2020robust}
T.~Takayama, A.~Yaresko, A.~Gibbs, K.~Ishii, D.~Kukusta, H.~Takagi,
\newblock \emph{Physical Review Materials} \textbf{2020}, \emph{4}, 7 075002.

\bibitem{bhandari1980silicon}
C.~Bhandari, D.~Rowe,
\newblock \emph{Contemporary physics} \textbf{1980}, \emph{21}, 3 219.

\bibitem{settipalli2022effect}
M.~Settipalli, V.~S. Proshchenko, S.~Neogi,
\newblock \emph{Journal of Materials Chemistry C} \textbf{2022}, \emph{10}, 19 7525.

\bibitem{PhysRevB.102.235144}
E.~Emmanouilidou, S.~Mardanya, J.~Xing, P.~V.~S. Reddy, A.~Agarwal, T.-R. Chang, N.~Ni,
\newblock \emph{Phys. Rev. B} \textbf{2020}, \emph{102} 235144.

\bibitem{PhysRevMaterials.5.L091202}
C.~A.~C. Garcia, D.~M. Nenno, G.~Varnavides, P.~Narang,
\newblock \emph{Phys. Rev. Mater.} \textbf{2021}, \emph{5} L091202.

\bibitem{yao2023large}
X.~Yao, J.~Gaudet, R.~Verma, D.~E. Graf, H.-Y. Yang, F.~Bahrami, R.~Zhang, A.~A. Aczel, S.~Subedi, D.~H. Torchinsky, et~al.,
\newblock \emph{Physical Review X} \textbf{2023}, \emph{13}, 1 011035.

\bibitem{PhysRevB.96.235128}
H.-Y. Yang, T.~Nummy, H.~Li, S.~Jaszewski, M.~Abramchuk, D.~S. Dessau, F.~Tafti,
\newblock \emph{Phys. Rev. B} \textbf{2017}, \emph{96} 235128.

\bibitem{gaudet2021weyl}
J.~Gaudet, H.-Y. Yang, S.~Baidya, B.~Lu, G.~Xu, Y.~Zhao, J.~A. Rodriguez-Rivera, C.~M. Hoffmann, D.~E. Graf, D.~H. Torchinsky, et~al.,
\newblock \emph{Nature materials} \textbf{2021}, \emph{20}, 12 1650.

\bibitem{tian2017understanding}
Y.~Tian, S.~Jia, R.~J. Cava, R.~Zhong, J.~Schneeloch, G.~Gu, K.~S. Burch,
\newblock \emph{Physical Review B} \textbf{2017}, \emph{95}, 9 094104.

\bibitem{yoon2011negative}
D.~Yoon, Y.-W. Son, H.~Cheong,
\newblock \emph{Nano letters} \textbf{2011}, \emph{11}, 8 3227.

\bibitem{riccardi2019probing}
E.~Riccardi, O.~Kashuba, M.~Cazayous, M.-A. M{\'e}asson, A.~Sacuto, Y.~Gallais,
\newblock \emph{Physical Review Materials} \textbf{2019}, \emph{3}, 1 014002.

\bibitem{PhysRevB.100.201107}
A.~Zhang, X.~Ma, C.~Liu, R.~Lou, Y.~Wang, Q.~Yu, Y.~Wang, T.-l. Xia, S.~Wang, L.~Zhang, X.~Wang, C.~Chen, Q.~Zhang,
\newblock \emph{Phys. Rev. B} \textbf{2019}, \emph{100} 201107.

\bibitem{shojaei2021raman}
I.~A. Shojaei, S.~Pournia, C.~Le, B.~R. Ortiz, G.~Jnawali, F.-C. Zhang, S.~D. Wilson, H.~E. Jackson, L.~M. Smith,
\newblock \emph{Scientific Reports} \textbf{2021}, \emph{11}, 1 8155.

\bibitem{wang2022axial}
Y.~Wang, I.~Petrides, G.~McNamara, M.~M. Hosen, S.~Lei, Y.-C. Wu, J.~L. Hart, H.~Lv, J.~Yan, D.~Xiao, et~al.,
\newblock \emph{Nature} \textbf{2022}, \emph{606}, 7916 896.

\bibitem{han1993anharmonic}
Y.-J. Han, P.~Klemens,
\newblock \emph{Physical Review B} \textbf{1993}, \emph{48}, 9 6033.

\bibitem{xu2017temperature}
B.~Xu, Y.~M. Dai, L.~X. Zhao, K.~Wang, R.~Yang, W.~Zhang, J.~Y. Liu, H.~Xiao, G.~Chen, S.~A. Trugman, et~al.,
\newblock \emph{Nature communications} \textbf{2017}, \emph{8}, 1 14933.

\bibitem{gall2021materials}
D.~Gall, J.~J. Cha, Z.~Chen, H.-J. Han, C.~Hinkle, J.~A. Robinson, R.~Sundararaman, R.~Torsi,
\newblock \emph{MRS Bulletin} \textbf{2021}, 1--8.

\bibitem{disa2021engineering}
A.~S. Disa, T.~F. Nova, A.~Cavalleri,
\newblock \emph{Nature Physics} \textbf{2021}, \emph{17}, 10 1087.

\bibitem{yu2023nontrivial}
J.~Yu, C.~J. Ciccarino, R.~Bianco, I.~Errea, P.~Narang, B.~A. Bernevig,
\newblock Nontrivial quantum geometry and the strength of electron-phonon coupling, \textbf{2023}.

\bibitem{garate2013phonon}
I.~Garate,
\newblock \emph{Physical Review Letters} \textbf{2013}, \emph{110}, 4 046402.

\bibitem{xu2022catalogue}
Y.~Xu, M.~G. Vergniory, D.-S. Ma, J.~L. Mañes, Z.-D. Song, B.~A. Bernevig, N.~Regnault, L.~Elcoro,
\newblock Catalogue of topological phonon materials, \textbf{2022}.

\bibitem{PhysRevResearch.4.043060}
F.~Nathan, I.~Martin, G.~Refael,
\newblock \emph{Phys. Rev. Res.} \textbf{2022}, \emph{4} 043060.

\bibitem{10.1063/1.5143061}
P.~Blaha, K.~Schwarz, F.~Tran, R.~Laskowski, G.~K.~H. Madsen, L.~D. Marks,
\newblock \emph{The Journal of Chemical Physics} \textbf{2020}, \emph{152}, 7 074101.

\bibitem{PhysRevLett.77.3865}
J.~P. Perdew, K.~Burke, M.~Ernzerhof,
\newblock \emph{Phys. Rev. Lett.} \textbf{1996}, \emph{77} 3865.

\bibitem{ROURKE2012324}
P.~Rourke, S.~Julian,
\newblock \emph{Computer Physics Communications} \textbf{2012}, \emph{183}, 2 324.

\bibitem{shoenberg2009magnetic}
D.~Shoenberg,
\newblock \emph{Magnetic Oscillations in Metals},
\newblock Cambridge Monographs on Physics. Cambridge University Press, \textbf{2009}.

\bibitem{sundararaman2017jdftx}
R.~Sundararaman, K.~Letchworth-Weaver, K.~A. Schwarz, D.~Gunceler, Y.~Ozhabes, T.~Arias,
\newblock \emph{SoftwareX} \textbf{2017}, \emph{6} 278.

\bibitem{dal2014pseudopotentials}
A.~Dal~Corso,
\newblock \emph{Computational Materials Science} \textbf{2014}, \emph{95} 337.

\bibitem{rappe1990optimized}
A.~M. Rappe, K.~M. Rabe, E.~Kaxiras, J.~Joannopoulos,
\newblock \emph{Physical Review B} \textbf{1990}, \emph{41}, 2 1227.

\bibitem{perdew2008density}
J.~P. Perdew, V.~N. Staroverov, J.~Tao, G.~E. Scuseria,
\newblock \emph{Physical Review A} \textbf{2008}, \emph{78}, 5 052513.

\bibitem{marzari1997maximally}
N.~Marzari, D.~Vanderbilt,
\newblock \emph{Physical review B} \textbf{1997}, \emph{56}, 20 12847.

\bibitem{souza2001maximally}
I.~Souza, N.~Marzari, D.~Vanderbilt,
\newblock \emph{Physical Review B} \textbf{2001}, \emph{65}, 3 035109.

\bibitem{giustino2007electron}
F.~Giustino, M.~L. Cohen, S.~G. Louie,
\newblock \emph{Physical Review B} \textbf{2007}, \emph{76}, 16 165108.

\end{thebibliography}

\newpage
\section*{Figure Legend}

\textbf{Figure 1:}
\noindent
\textbf{Anomalous Transport in Various Metals} a) Transport and single particle mobilities for various compounds hosting strong phonon-electron coupling. These materials all exhibit orders of magnitude difference between their single particle and bulk transport mobilities despite significant differences in topology, symmetry, and bulk band structures.  b) Cartoon schematic of electron scattering pathways in a normal metal and a phonon-electron fluid. The faucet represents an electric field providing a "bath" of electrons. The momentum of these electrons can either be transferred to defect scattering, or into the phonon bath. From here, with strong phonon-phonon coupling the momentum is lost to anharmonic phonon decay and heat. For the case of strong phonon-electron coupling the momentum is returned to the electron bath, greatly enhancing bulk transport even with relatively low single particle mobilities.

\noindent
\textbf{Figure 2:}
\noindent
\textbf{Electronic Properties of MX$_2$} ) Single particle and transport mobility of all four compounds in the MX$_2$ family. As the series evolves from NbGe$_2$ to TaSi$_2$, a clear trend emerges with the transport mobility ($\mu_T$) decreasing dramatically despite larger single particle mobilities ($\mu_Q$). b) The experimental ($m^*_{exp}$) and theoretical ($m^*_{theory}$) quasiparticle effective masses plotted as a function of dHvA frequencies. A substantial enhancement in the experimental masses over the DFT calculated ones is observed in the MGe$_2$, but not MSi$_2$. Nonetheless, the overall change in mass between the materials does not explain the trends in mobilities.

\noindent
\textbf{Figure 3:}
\noindent
\textbf{Phonon Linewidths} Temperature dependence of the Raman linewidths for A$_1$ and E$_2$ modes in a) MSi$_2$ and b) MGe$_2$. 
    Measured data are shown in black, the blue curve is a fit to the phonon-electron decay model, while the red curve is the typical anharmonic phonon-phonon contribution. The black line is a sum of both contributions and gives the total fit. We see clear evidence of phonon-phonon dominated linewidths in \ch{MSi2} and phonon-electron dominated linewidths in \ch{MGe2}.

\noindent
\textbf{Figure 4:}
\noindent
\textbf{Transport and Linewidths} a)Cartoon schematic of anharmonic decay of an optical phonon into a pair of acoustic phonons, conserving energy and momentum.  b) Schematics of phonon decay into an electron-hole pair plotted on the bandstructure of \ch{NbGe2}. $\omega_0$ is the phonon energy, the orange arrow is $\omega_A$ and represents the separation of the band from the Fermi level. The blue arrow represents an electron-hole excitation of the same energy as the incident phonon.
    c) Phonon-electron and anharmonic contributions to Raman linewidths of every phonon mode in NbGe$_2$, TaGe$_2$, NbSi$_2$ and TaSi$_2$. The phonon-electron contribution heavily dominates the \ch{MGe2} linewidths, while the phonon-phonon contribution dominates the \ch{MSi2} linewidths.

\noindent
\textbf{Figure 5:}
\noindent
\textbf{DFT Dispersions and JDOS} a) Calculated vs measured phonon frequencies for \ch{NbGe2} and \ch{NbSi2}. The black line indicates the ideal agreement. b) Phonon-phonon joint density of states (JDOS) for NbGe$_2$ and NbSi$_2$ which show nearly equal spectral weight despite reduced acoustic bandwidth in \ch{NbSi2} c) Phonon-electron JDOS for NbGe$_2$ and NbSi$_2$ which shows approximately 20\% larger phase space for phonon-electron decay in NbGe$_2$ than NbSi$_2$. d) Phonon dispersions of NbGe$_2$ and NbSi$_2$ are shown in blue and red, respectively. Dispersions show approximately double the acoustic bandwidth for NbSi$_2$ than NbGe$_2$. e,) Calculated phonon-phonon lifetimes at 50K for \ch{NbGe2} and \ch{NbSi2}, respectively.

\section*{Methods}
\subsection{Crystal growth}
Single crystals of MX$_2$ were synthesized by chemical vapor transport (CVT) technique with iodine as the transport agent. 
The starting elements were all high-purity powders, mixed in stoichiometric ratios with 50 mg of iodine, sealed in silica tubes under vacuum. 
We found the best conditions to make high-quality crystals were to place the tubes inside a box furnace (minimal temperature gradient) and keep it at 900$^{\circ}$C for one month.
\subsection{Electrical transport measurement}
Resistivity and Hall effect were measured using a standard four-probe technique using the Quantum Design PPMS Dynacool instrument. 
The heat capacity was measured using PPMS with a relaxation time method on a piece of poly-crystalline sample cut from sintered pellets.
\subsection{Quantum oscillation}
Quantum oscillation experiments under continuous fields up to 41 T were performed at the National High Magnetic Field Laboratory in Tallahassee, Florida. 
A piezo-resistive cantilever technique was used to measure the dHvA effect (Piezo-resistive self-sensing 300 $\times$ 100 $\mu$m cantilever probe, SCL-Sensor. Tech.). 
A $^3$He cryostat was used for high-field experiments at temperature down to 0.3 K.
\subsection{Density functional theory}
Density functional theory (DFT) calculations with the full-potential linearized augmented plane-wave (LAPW) method were implemented in the WIEN2k code~\cite{10.1063/1.5143061} using the Perdew-Burke-Ernzerhof (PBE) exchange-correlation functional~\cite{PhysRevLett.77.3865} and spin-orbit coupling (SOC). We set the basis-size control parameter to RK$_{max}$=8.5, and use 20000 k-points to sample the k-space. We obtain fermi surface from Xcrysden, combined with our DFT band structure calculations. The Supercell K-space Extremal Area Finder (SKEAF) program~\cite{ROURKE2012324} was applied to find dHvA frequencies and effective masses of different Fermi pockets \cite{shoenberg2009magnetic,PhysRevB.96.235128}.

\subsection{Raman Scattering}
 All data was taken from 10~K - 300~K using a 532nm laser in a backscattering configuration. For each of the samples that were measured, 3 spectra were taken at each temperature to allow for the removal of cosmic spikes and the data was averaged to provide a single spectrum for each temperature. To extract the temperature dependent behavior of the phonons, each phonon mode is fit to a Voigt profile. The resulting energies and linewidths are then fit to our combined models for phonon-electron and anharmonic scattering.
 
\subsection{Computational Details}
For the weighted joint density of states (JDOS) and refractive index evaluations, separate DFT calculations were carried out using the implementation in JDFTx~\cite{sundararaman2017jdftx}. Fully relativistic ultrasoft pseudo-potentials~\cite{dal2014pseudopotentials,rappe1990optimized} for the PBEsol exchange-correlation functional~\cite{perdew2008density} were used, as well as a uniform $6\times 6\times 8$ $k$ grid for the 6-atom standard primitive unit cell, an energy cutoff of 28 Hartrees, Fermi-Dirac smearing with a 0.01 Hartree width, and a $3\times 3\times 2$ phonon supercell. Maximally localized Wannier functions~\cite{marzari1997maximally,souza2001maximally} were similarly obtained to interpolate quantities for Monte Carlo Brillouin zone integration on finer $k$ and $q$ meshes.~\cite{giustino2007electron} Each transition contributing to the JDOS at that transition energy is weighted by the factor $n_F(E_{\mathbf{k},n},T)-n_F(E_{\mathbf{k},m},T)$, where $n_F(E,T)$ is the Fermi-Dirac distribution function, and the transition occurs between energies $E_{\mathbf{k},n}$ and $E_{\mathbf{k},m}$ at the momentum $\mathbf{k}$ for two bands indexed by $n$ and $m$.

\section*{Acknowledgement}
V.M.P., Yiping Wang, B.S., M.R., G.M., and K.S.B. acknowledge the primary support of the US Department of Energy (DOE), Office of Science, Office of Basic Energy Sciences under award number DE-SC0018675. F.T. and X.Y. acknowledge support from the US Department of Energy, Office of Basic Energy Sciences, Division of Physical Behavior of Materials under Award No. DE-SC0023124. Yaxian Wang acknowledges funding support from the Chinese Academy of Sciences (Nos.~YSBR047 and E2K5071). J. Y. C. and G.T.M acknowledge support from the National Science Foundation under the award number NSF/DMR-2209804 and the Welch Foundation under the award number AA-2056-20220101. The National High Magnetic Field Laboratory is supported by the National Science Foundation through NSF/DMR-1644779 and NSF/DMR-2128556 and by the State of Florida. 

\section{Author contributions:} V.M.P. performed the Raman experiments and analyzed the data. B.S. and G.M. assisted in the Raman data collection. M.R. and Y.W. assisted in the setup of the automated low temperature microscope. F.T., H.Y. grew the \ch{MX2} crystals. G.T.M. and J.Y.C. performed the temperature dependent X-ray diffraction experiments. X.Y. performed the Hall measurements and analyzed them along with the Quantum oscillations. A.S., D.G., and E.S.C performed the de Haas - van Alphen measurements. P.N., Y.W. and G.V. performed first-principles calculations. V.M.P and X.Y. wrote the manuscript with the help of K.S.B., F.T., and P.N; and with contributions from authors. K.S.B. conceived and supervised the project. 

\section{Competing interests:} The authors declare no competing interests. 

\section{Data and materials availability:} All data are available in the main text or the supplementary materials. The data supporting the plots within this paper and other study findings are available from the corresponding authors on request.

\end{document}